\documentclass[12pt,titlepage]{article}
\usepackage{epsfig}
\def\L{{\bf L}}
\def\F{{\bf F}}

\def\G2{{\Gamma(2)}}

\def\beq{\begin{equation}}
\def\eeq{\end{equation}}
\begin{document}
\title{Modular symmetry and 
temperature flow of conductivities
in quantum Hall systems with varying Zeeman energy}
\author{{Brian P. Dolan}\\
{\small Department of Mathematical Physics,}\\
{\small National University
of Ireland, Maynooth, Ireland}\\
{\small and}\\
{\small Dublin Institute for Advanced Studies,
10 Burlington Rd, Dublin, Ireland}\\
{\small e-mail: bdolan@thphys.nuim.ie}}
\maketitle
\begin{abstract}

The behaviour of the critical point between quantum Hall plateaux,
as the Zeeman energy is varied, is analysed using modular symmetry
of the Hall conductivities following from the law of corresponding states.
Flow diagrams for the conductivities as a function of temperature,
with the magnetic field fixed, are constructed for different Zeeman
energies, for samples with particle-hole symmetry.

\bigskip

\noindent\hbox{Preprint no: DIAS-STP-10-08}

\medskip

\noindent\hbox{PACS nos: 73.43.Nq, 05.30.Fk, 05.30.Rt, 02.20.-a}

\end{abstract}

\section{Introduction}

The quantum Hall effect continuous to intrigue both experimentalists
and theorists not only because of the beautifully rich patterns visible
in the data but also because of the fascinating physics involved in
the collective phenomena of strongly interacting systems.
The first suggestion of a connection between the quantum Hall effect
and the modular group appeared in \cite{Shapere+Wilczek}, although
these authors focused on a sub-group of the full modular group that did not
turn out to have any direct relevance to the current experimental data.
Subsequent papers on symmetries
of the phase diagram of the quantum Hall effect 
\cite{KLZ,Lutken+Ross} appeared almost simultaneously from two very different
directions and laid the foundations for the application of
modular symmetry to the quantum Hall effect.  Although reference \cite{KLZ}
did not use the mathematical language of modular symmetry the \lq\lq Law
of Corresponding States'' put forward in that reference is in fact equivalent
to the assumption of modular symmetry \cite{BDa}.
 
Modular symmetry gives predictions \cite{BDc} for the manner in which
the conductivity of a two-dimensional quantum Hall sample flows,
as the temperature is varied keeping the magnetic field fixed ---
predictions which have already received strong experimental 
support \cite{Murzin1,Murzin2,Taiwan1,Taiwan2,Taiwan3}.

While the flow diagram presented in \cite{BDc}
was for spin-split samples the experimental data 
presented in \cite{Murzin1} is for spin-degenerate samples.
Zeeman splitting in the context of modular symmetry was analysed 
in \cite{BDspin}
and a flow diagram, deformed by interactions between adjacent Landau levels,
was given but 
the fully spin-degenerate case has not yet been treated using modular symmetry

In this paper restrictions on the temperature flow due to modular symmetry are 
combined with a Zeeman splitting analysis 
to determine the way in which 
temperature flow changes as the Zeeman splitting is smoothly varied
between the two extremes of samples with well-split spins and fully spin-degenerate
samples.  The analysis is restricted to situations with particle-hole symmetry,
as modular symmetry is particularly powerful in this case but does not give
strong predictions otherwise.
A central prediction of the analysis is that pairs of 
critical points of the quantum Hall phase transitions
between adjacent plateaux in spin-split samples must merge as the Zeeman splitting
is reduced, as shown in figures 12 and 13.

\section{The modular group}

The law of corresponding states \cite{KLZ,ZHK}, for isotropic
quantum Hall samples with
spins well split, can be written in terms of the complex 
conductivity 
\[ \sigma = \sigma_{xy} + i\sigma_{xx}\]
(for isotropic samples $\sigma_{xx}=\sigma_{yy}$).
A general map between two quantum Hall states can be constructed by iterating
two generating maps, \cite{Lutken+Ross,BDa,BL}: 
the Landau level addition transformation, ${\bf L}$,
\[ \label{Complexsigma}
\sigma\rightarrow \sigma + 1,
\]
and the flux attachment transformation, ${\bf F}^2$,
\[
\sigma\rightarrow \frac{\sigma}{2\sigma+1},
\]
which attaches two units of statistical gauge field flux to each
electron (we use units with $\frac{e^2}{h}=1$).

In samples which enjoy particle-hole symmetry there is a third
map, the particle-hole transformation, ${\bf P}_1$,
$$
\sigma\rightarrow 1 - \overline\sigma.
$$
These maps generalise Jain's transformations on ground state wavefunctions,
\cite{Jain}, to include non-zero Ohmic conductivity.
Repeated iteration of ${\bf F}^2$ and ${\bf L}$ generate an infinite discrete
group which we shall denote $\Gamma_0(2)$.

The infinite discrete group generated by repeated applications of
${\bf L}$ and ${\bf F}$ is called the modular group and is usually
denoted by $\Gamma(1)$ in the mathematical literature \cite{Koblitz}. 
A general element $\gamma\in\Gamma(1)$
sends
\begin{equation}
\label{gammadef}
\sigma\ \rightarrow\ \gamma(\sigma):=\frac{a\sigma +b}{c\sigma +d},
\end{equation}
where $a,b,c$ and $d$ are any four integers satisfying $ad-bc =1$.
Group multiplication can be realised in terms of the $2\times 2$ matrix
\begin{equation}\label{gammaM}
\gamma=\left(
\begin{array}{cc}a&b\\ c&d\\ \end{array}
\right)\end{equation}
and demanding that $\det\gamma=1$. It is easy to check from the definition
(\ref{gammadef}) that, for any three
such matrices satisfying $\gamma_1\gamma_2=\gamma_3$, 
we have $\gamma_1(\gamma_2(\sigma))=\gamma_3(\sigma)$. Thus the group multiplication
law is given by matrix multiplication.

The full modular group $\Gamma(1)$ is not a symmetry of quantum Hall effect, for example
the element $\left(\begin{array}{cc} 0&1\\ -1&0\\ \end{array}\right)$
sends $\sigma \rightarrow -1/\sigma$ which has a fixed point for $\sigma=i$,
{\it i.e.} $\sigma_{xx}=1$, $\sigma_{xy}=0$.
There is no indication in the experimental data on the quantum
Hall effect that this point has any special significance (though it is
important in the insulator --- superconductor phase transition \cite{MFisher,Gammatheta}).

The group $\Gamma_0(2)$ is a sub-group of the modular group, it is represented
by matrices of the from (\ref{gammaM}) with the extra condition that $c$ be even.
In matrix notation Landau level addition and Flux attachment are represented
by
$\L=
\left(
\begin{array}{cc} 1&1\\ 0&1 \\ \end{array}
\right)$ and $\F^2=
\left(\begin{array}{cc} 1&0\\ 2&1\\ \end{array}
\right).$

If the electron spins are not split then the story is different.
For simplicity first consider the situation if Zeeman splitting is completely
absent and states corresponding to two different electron spins
are completely degenerate.  Then Landau level addition sends
\[ \sigma\rightarrow \sigma + 2, \]
or $\frac\sigma 2 \rightarrow \frac \sigma 2 +1$,
which is ${\bf L}^2=\left(\begin{array}{cc} 1 & 2 \\ 0 & 1 \\ \end{array}\right)$.
Attaching two units of flux to each electron sends
\[ 
\frac{\sigma} {2} \rightarrow  
\frac {\left(\frac{\sigma} {2}\right)} {2\left(\frac{\sigma} {2}\right) + 1}
\qquad \Rightarrow \qquad \sigma \rightarrow \frac {\sigma}{\sigma + 1},\]
which is $\F=\left(\begin{array}{cc} 1 & 0 \\ 1 & 1 \\ \end{array}\right)$ acting
on $\sigma$.  Thus spin generate Landau levels give rise to an infinite discrete
group generate by $\L^2$ and $\F$.  A general element of this group  
is of the form $\gamma=\left(\begin{array}{cc} a & b \\ c & d \\ \end{array}\right)$
with $a$, $b$, $c$ and $d$ integers satisfying $ad-bc=1$,
but with $b$ restricted to be even.
We denote this group by $\Gamma^0(2)$.  In fact
that $\Gamma^0(2)$  acting on $\sigma$ is the same as $\Gamma_0(2)$
acting on $\frac \sigma 2$.  Particle-hole symmetry for degenerate
spins is also modified to ${\bf P}_2=\sigma\rightarrow 2-\overline\sigma$,
so ${\bf P}_2={\bf P}_1+1$.

For spins that have a slight splitting, so that the Landau levels 
corresponding to opposite spins are not
completely degenerate but are still close enough for there to be some
mixing between Landau level wave-functions, one expects $\L^2$ to be a symmetry rather
than $\L$.  At the same time adding two units of statistical flux to 
each electron in the individual levels is described by $\F^2$.
For this intermediate case therefore the group is generated by
$\L^2$ and $\F^2$ and a general element is represented by
$\gamma=\left(\begin{array}{cc} a & b \\ c & d \\ \end{array}\right)$
with $ad-bc=1$ and {\it both} $b$ and $c$ even integers. 
This group is denoted by $\Gamma(2)$.
Clearly $\Gamma(2)\subset \Gamma_0(2)$ and  
$\Gamma(2)\subset \Gamma^0(2)$.
We are led to suggest the following sequence of symmetries as the Zeeman
splitting relative to the cyclotron energy is varied from large to small:
\[\Gamma_0(2)\quad\longrightarrow\quad\Gamma(2)\quad\longrightarrow\quad
\Gamma^0(2),\]
thus the symmetry first decreases and then increases again
as the Zeeman splitting is varied.

Since the modular group has an infinite number of elements the law
of corresponding states maps between an infinite number of quantum Hall
phases, which is clearly a mathematical idealisation which is never
realised in any physical system.  Obviously the symmetry breaks down
in various limits, such as weak magnetic field, when the Landau level
splitting becomes comparable with thermal energies, or very strong
magnetic fields, when a Wigner crystal is expected to form.
The range of validity is discussed in \cite{BDb}.  

It was pointed out in \cite{Sachdev} that the law of corresponding states
applies to the AC conductivity in the limit of infinite frequency
rather than the DC conductivity.  This is because the conductivity is a
a function of frequency over temperature, 
$\sigma_{xx}\left(\frac \omega T\right)$, 
and the derivation of the law of corresponding states in \cite{KLZ}
takes the limit $T\rightarrow 0$ before $\omega\rightarrow 0$ 
and this does
not in general commute with the limit required to extract the DC conductivity,
namely $\omega\rightarrow 0$ before $T\rightarrow 0$.
The precise relationship between these two
limits can only be explored in the context of a specific microscopic
theory for the conductivity and is not accessible solely through infra-red
effective action techniques such as the law of corresponding states.  
Nevertheless the law of corresponding states
has been applied to DC conductivities, for example with regard to 
temperature flows, and the experimental data are in 
remarkable agreement with the predictions 
\cite{Murzin1,Murzin2,Taiwan1,Taiwan2,Taiwan3,BDb}.  
While some microscopic models may display a
symmetry which makes the order in which the limits are taken
irrelevant  \cite{Sachdev}
this is not generic, but since we do not commit ourselves to a specific
model here we cannot address this question directly.  Rather our philosophy will
be to develop the predictions of the law of corresponding states, see
where they lead, and future experiments will test their validity.

\section{Temperature flow}

Scaling arguments \cite{Wei} suggest that, at low temperatures, the
DC conductivity should be a function of a single variable,
$\sigma\left(\frac{\Delta B} {T^\kappa}\right)$,
rather than of the temperature $T$ and magnetic field $B$
separately.
Here $\Delta B= B- B_c$ is the deviation of the magnetic field
from the critical value $B_c$ separating two quantum Hall
phases and $\kappa$ is a scaling exponent, 
experimentally $\kappa\approx 0.44$.
The conductivity depends on 
the electron scattering length $l$ and we define the scaling
function 
\[\Sigma_l(\sigma,\overline\sigma)=-l \frac {d\sigma}{dl}.\]
Assuming $l$ is a strictly monotonic function of temperature, increasing as
$T$ decreases, the temperature flow described by the scaling function
\[\Sigma_T(\sigma,\overline\sigma)= T \frac {d\sigma}{d T}\]
will have the same topology as the flow described by $\Sigma_l$ ---
they will have the same fixed points, $\Sigma_T=0$ if and only if 
$\Sigma_l=0$.  We do not need to determine either $\Sigma_l$
or $\Sigma_T$ exactly, in fact if $s(T)$ is {\it any} monotonic
function of $T$, decreasing as $T$ decreases, then
the flow described by
\[ \Sigma_s(\sigma,\overline \sigma)=  s \frac {d\sigma}{d s} \]
will have the same topology as both that of $\Sigma_T$ and of $\Sigma_l$.

If the law of corresponding states 
correctly describes the low temperature 
physics then the scaling flow commutes with the
law of corresponding states map.  From this it can be concluded that
any value of the complex conductivity $\sigma_*$ 
that is a fixed point of the modular group
(in the sense that there exists an element $\gamma$ of the modular group
such that $\gamma(\sigma_*)=\sigma_*$)
must also be a fixed point of the scaling flow, \cite{BDa,BL}.
This implies that
\[\Sigma_s(\sigma_*,\overline \sigma_*)=0.\]
This follows because the assumption $\gamma(\sigma_*)=\sigma_*$ requires
\beq\label{SigmaFlow}
\Sigma_s(\sigma_*,\overline \sigma_*)=\Sigma_s\bigl(\gamma(\sigma_*),\gamma(\overline\sigma_*)\bigr)=s\left.\frac {d \gamma(\sigma)}{d s}\right|_{\sigma_*}
=\frac 1 {(c\sigma_* + d)^2}\Sigma_s(\sigma_*,\overline \sigma_*)\eeq
which is only possible if $(c\sigma_* +d)^2=1$ or 
if $\Sigma_s(\sigma_*,\overline \sigma_*)=0$ or $\infty$.
If $(c\sigma_* + d)^2=1$ then
$a\sigma_* +b =\pm \sigma_*$ and, excluding the trivial case $a=\pm 1$, $b=0$,
this is 
not possible if the Ohmic conductivity $(\sigma_*)_{xx}>0$ at the fixed point.
Assuming $\Sigma_s(\sigma_*,\overline\sigma_*)$ is not infinite
we conclude that 
$\Sigma_s(\sigma_*,\overline \sigma_*)=0$ 
and $\sigma_*$ is a fixed point of the flow.
The fixed points with  $Im(\sigma_*)>0$ are isolated and easy to enumerate, 
since
\[\gamma(\sigma_*)=\sigma_* \qquad \Rightarrow 
\qquad \sigma_*=\frac{a-d \pm\sqrt{(a+d)^2-4}} {2c}.\]
Now $ad-bc=1$, with $bc$ even for all three groups $\Gamma_0(2)$, $\Gamma^0(2)$
and $\Gamma(2)$, hence $ad$ must be odd so both $a$ and $d$
must be odd.  Demanding $Im(\sigma_*)>0$ then requires that $-2<a+d<2$.
Hence $a+d=0$, $\pm 1$, but $a$ and $d$ are both odd 
so $\pm 1$ is ruled out and we can conclude that $d=-a$.
Hence
\beq\label{sigmastar}
\sigma_*=\frac{a + i} {c}\eeq
as the Ohmic conductivity, $Im(\sigma_*)$, cannot be negative.
A matrix $\gamma$ that leaves $\sigma_*$ fixed must now be of the form
\[\gamma=\left(\begin{array}{cc} a & b \\ c & -a \\ \end{array}\right)\]
with $bc=-(1+a^2)$ with $a$ odd.  Let $a=2p+1$ for some integer $p$,
then $bc=-4p(p+1)-2$ and $bc=2$ mod $4$.  In particular $b$ and $c$ cannot
both be even so $\Gamma(2)$ has {\it no} fixed points with
$Im(\sigma)>0$.

To summarise: $\Gamma_0(2)$ has fixed points above the real axis
of the form (\ref{sigmastar}) with $a$ odd and $c$ even;
$\Gamma^0(2)$ has fixed points above the real axis
of the form (\ref{sigmastar}) with $a$ and $c$ both odd;
$\Gamma(2)$ has no fixed points above the real axis.

Although any fixed point of the modular group with $Im(\sigma_*)>0$ must be
a fixed point of the flow the converse does not necessarily hold,
there could be fixed points of the flow that are not fixed points of
the modular group.  Any such point would have an infinite number of images
under the group action.  However there is no
sign any such extra fixed points in the experimental data for spin split 
samples so we shall assume that there are no fixed points 
for $\Gamma_0(2)$, other than those required by the symmetry.
For  brevity in the following this will be referred as the 
{\it minimalist assumption}.
If we assume that the topology of the flow varies smoothly as the
Zeeman energy is varied then the $\Gamma_0(2)$ fixed points cannot
suddenly disappear when $\Gamma_0(2)$ is broken to the smaller group
$\Gamma(2)$, they must move down towards the real axis.
We shall likewise assume that the only fixed points of the flow for
degenerate spins are those of $\Gamma^0(2)$ and these move smoothly down
towards the real axis as the degeneracy is lifted.

The topology of the flow is determined by the fixed points and
some other rather mild assumptions \cite{BDc}, 
such as decreasing Ohmic conductivity when $\sigma_{xx}>>\sigma_{xy}$ as the temperature is reduced, as in a semi-conductor, and attractive fixed
points at integer quantum Hall plateaux.    
We can plot the flow by changing variables from $\sigma$
to $\lambda(\sigma)$ where $\lambda$ is invariant under $\Gamma(2)$, {\it i.e.}
$\lambda\bigl(\gamma(\sigma)\bigr)=\lambda(\sigma)$, with $\gamma\in\Gamma(2)$,
\cite{BDspin,SemiCircle}.  Since $\Sigma_s(\sigma,\overline\sigma)$ represents a vector flow in
a two-dimensional space this is just a change of co-ordinates in that
space.  In the new parametrisation the flow is given by
\[
\Sigma_\lambda(\lambda,\overline\lambda)=s\frac{d \lambda}{d s}
=\Sigma_s(\sigma,\overline\sigma)\lambda'
\qquad \Rightarrow \qquad \Sigma_s(\sigma,\overline\sigma)=\frac{\Sigma_\lambda}{\lambda'} ,\]
where $\lambda'=\frac{d\lambda}{d\sigma}$.

The invariant function $\lambda$ that has the smallest number of poles
and zeros in the complex plane is unique, up to
a constant rescaling and addition of a constant, \cite{WW}.  
It is most easily expressed in terms of Jacobi
$\vartheta$-functions: 
\[\lambda=\frac{\vartheta_2^4}{\vartheta_3^4}\]
where
\beq \vartheta_3(\sigma)=\sum_{n=-\infty}^\infty e^{i\pi n^2\sigma}\qquad
\hbox{and}\qquad
\vartheta_2=2\sum_{n=0}^\infty e^{i\pi(n+\frac{1}{2})^2\sigma}.\eeq
It is also useful to define
\beq \label{theta-identity}
\vartheta^4_4(\sigma)=\sum_{n=-\infty}^\infty (-1)^n e^{i\pi n^2\sigma}
=\vartheta^4_3(\sigma)-\vartheta^4_2(\sigma)
.\eeq
Then these functions have the following transformation properties
under ${\bf L}$ and ${\bf F}^2$:
\begin{eqnarray}
\vartheta_2(\sigma+1)=e^{\frac{i\pi} {4}}\vartheta_2(\sigma) ,\kern -12pt
&\qquad & \vartheta_2\left(\frac{\sigma}{2\sigma+1}\right)=\sqrt{2\sigma+1}\;\vartheta_2(\sigma),\nonumber \\
\vartheta_3(\sigma+1)=\vartheta_4(\sigma),
\;&\quad\qquad & \vartheta_3\left(\frac{\sigma}{2\sigma+1}\right)=\sqrt{2\sigma+1}\;\vartheta_3(\sigma),\nonumber \\
\vartheta_4(\sigma+1)=\vartheta_3(\sigma), 
\;&\quad\qquad & \vartheta_4\left(\frac{\sigma}{2\sigma+1}\right)=
\sqrt{2\sigma+1}\;e^{-\frac{i\pi}{2}}\vartheta_4(\sigma),\nonumber
\end{eqnarray}
(we use the notation of \cite{WW}).

Furthermore under ${\bf P}_2$, $\vartheta_i \rightarrow \overline\vartheta_i$
for $i=2,3,4$.  Hence particle-hole interchange swaps $\lambda\leftrightarrow 
\overline\lambda$ and assuming particle-hole symmetry
has the important consequence for the scaling function that
\beq\label{PHinterchange}
\Sigma_\lambda(\lambda,\overline\lambda)=
\overline{\Sigma_\lambda(\lambda,\overline\lambda)}
=\Sigma_\lambda(\overline\lambda,\lambda).\eeq
This then implies that an expansion of $\Sigma$ in powers of $\lambda$ 
and $\overline\lambda$ has only real co-efficients and we draw
the important conclusion that, starting from
any point for which $\lambda$ is real, the flow 
can never generate an imaginary part for $\lambda$.  In other words
any curve on which $\lambda$ is real is in integral curve of the
flow \cite{SemiCircle}: this is a key observation in creating the
flow diagrams below.  In particular $\lambda$ is real on vertical
lines above the integer points on the real line and one the semi-circles
of radius $1/2$ joining the integers (see figure 1).

While $\lambda$ is not invariant under the larger groups $\Gamma_0(2)$ and
$\Gamma^0(2)$ the following functions of $\lambda$ are:
\begin{itemize}
\item
$\mu=\frac{\lambda-1}{\lambda^2}=
-\frac{\vartheta_3^4\vartheta_4^4}{\vartheta_2^8}$ is invariant under $\Gamma_0(2)$;\hfill\break
\item
$\rho=\frac{\lambda}{(1-\lambda)^2}=
\frac{\vartheta_2^4\vartheta_3^4}{\vartheta_4^8}$ is invariant under $\Gamma^0(2)$.  
\end{itemize}

The functions $\mu$ and $\rho$ can be used to determine the topology
of the flow for $\Gamma_0(2)$ and $\Gamma^0(2)$ respectively.
Fixed points of $\Gamma_0(2)$, respectively $\Gamma^0(2)$,
must be fixed points of the flow, and these are enumerated in equation
(\ref{sigmastar}).
Next, as above, we argue that demanding particle-hole symmetry
implies that curves on which $\mu$, respectively $\rho$, is real
will be integral curves of the flow.
The flow can then be modelled qualitatively
by taking $\Sigma_s$ to be a meromorphic function of $\sigma$, $\Sigma_s(\sigma)$.  Although
there is no physical argument for meromorphicity by plotting the flow
in this case we can get a picture of what it looks like, the inclusion 
of $\overline\sigma$ dependence can  only result in a smooth deformation
of the meromorphic flow which leaves the fixed points invariant.
Meromorphic functions satisfying (\ref{SigmaFlow}) are called modular
forms in the mathematical literature \cite{Koblitz} and their properties
are well known.
For $\Gamma_0(2)$ it 
was shown in \cite{BDc} that the minimalist assumption leads to
\beq\label{split-flow}
\Sigma_s(\sigma)=-\frac{\mu}{\mu'}=
\frac{\lambda(1-\lambda)}{\lambda'(2-\lambda)}=
\frac{1}{i\pi(2\vartheta_3^4 - \vartheta_2^4)},\eeq
where $\mu'=\frac{d\mu}{d\sigma}$ (the second and third forms of this
equation rely on equation (\ref{theta-identity}) and the fact that
\beq \label{lambdaprime}\frac{\lambda(1-\lambda)}{\lambda'}= \frac {1} {i\pi\vartheta_3^4},\eeq 
a result that can be derived using the techniques in \cite{BDc,BDspin}).  The integral curves of the 
flow (\ref{split-flow}) are plotted in figure 3.  

It should be borne in mind
that (\ref{split-flow}) does not give a quantitative description
of the temperature flow, because the function $s(T)$ is undetermined.
By changing $s(T)$, $\Sigma_s$ can be multiplied by a function of $T$,
but as long as $s(T)$ is strictly monotonic and real this will not
change the fixed points nor will it change the fact that curves on which
$\mu$ is real are integral curves of the flow --- it will merely change
the {\it rate} at which the flow lines in figure 3 are traversed
as the temperature is changed, and this rate is not evident in the
figure.  Also physical quantum Hall samples cannot be expected to 
give meromorphic flow, in a real sample $\Sigma(\sigma,\overline\sigma)$
will depend on $\overline\sigma$ and $\sigma$ independently, but 
similar arguments apply: $\overline\sigma$ dependence can only distort the
flow smoothly from the meromorphic flow shown  leaving the fixed points
invariant and, again assuming particle-hole symmetry, the vertical lines
above the integers and the semi-circles, together with their images
under $\Gamma_0(2)$, will not be affected (equation (\ref{PHinterchange})
does not require meromorphicity).  Figure 3 should be compared
to the experimental data in \cite{Murzin2,Taiwan1,Taiwan2} for
spin split samples.

For $\Gamma^0(2)$ similar arguments applied to $\rho$ lead to
\beq \label{degenerate-flow}
\Sigma_s(\sigma)= \frac{\rho}{\rho'}=\frac{\lambda(1-\lambda)}{\lambda'(1+\lambda)}
=\frac{1}{i\pi (\vartheta_3^4 + \vartheta_2^4)}\eeq
and this flow is plotted in figure 11.
This should be compared to the experimental flow for the
spin degenerate sample in \cite{Murzin1} -- the agreement is remarkable.

For samples intermediate between degenerate and well split spins
$\Gamma(2)$ symmetry is not as powerful as there are no fixed points
with $\sigma_{xx}>0$.  Nevertheless we would expect there to be fixed
points of the flow, the fixed points of $\Gamma_0(2)$ for spin split
samples and of $\Gamma^0(2)$ for spin degenerate samples can hardly just
disappear when the Zeeman splitting is smoothly varied.  
The minimalist assumption cannot be used for $\Gamma(2)$.
We can however assume that
the fixed points of $\Gamma_0(2)$ and $\Gamma^0(2)$
persist when the Zeeman splitting is varied, but
their position is no longer dictated by modular symmetry.  
We seek a smooth deformation from figure 3 to figure 11,
equation (\ref{split-flow}) to equation (\ref{degenerate-flow}),
as the Zeeman splitting is increased, a deformation which is compatible with
particle-hole symmetry, 
$\overline{\Sigma_\lambda(\lambda)}=\Sigma_\lambda(\overline\lambda)$.
In order to avoid creating new spurious fixed points we keep the
order of the polynomials in $\lambda$ fixed in the numerators and denominators
of (\ref{split-flow}) and (\ref{degenerate-flow}), and this
dictates that the interpolating  flow must be of the form
\beq\label{general-flowA}
\Sigma_s(\sigma)=
\frac{\lambda(A+B\lambda)}{\lambda'(C+D\lambda)}
,\eeq
where $A$, $B$, $C$ and $D$ are constants. 
Particle-hole symmetry requires $\overline {\Sigma_s(\sigma)}
= \Sigma_s(\overline{\sigma})$, which
dictates that $A$, $B$, $C$ and $D$ be real.  
Equation (\ref{lambdaprime}) shows that a factor $\frac{(A+B\lambda)}{\vartheta^4_3(1-\lambda)}$
will appear in $\Sigma_s(\sigma)$, unless $A=-B$, and this would
introduce a new zero when $\lambda=-\frac A B$ that is not there in
either (\ref{split-flow}) or (\ref{degenerate-flow}).
To avoid this we set $A=-B$ and the only possible deformation
that is compatible with our assumptions is, up to an overall constant factor,
\beq\label{general-flow}
\Sigma_s(\sigma)=
z\frac{\lambda(1-\lambda)}{\lambda'(\lambda+z)}=\frac{1}{i\pi}
\frac {z} {(\vartheta_2^4 + z \vartheta_3^4)},\eeq
with $z$ independent of $\sigma$ and real.  The free parameter $z$
varies from $z=- 2 $ for
$\Gamma_0(2)$ to $z=1$ for $\Gamma^0(2)$ (an overall factor of 2 multiplying
(\ref{split-flow}) does not change figure 3, indeed 
(\ref{split-flow}) is only derived in \cite{BDc} 
up to an overall positive constant). 

Figure 2 shows how the fixed point $\sigma_*=\frac{1+i}{2}$ of $\Gamma_0(2)$
moves as $z$ is varied: particle-hole symmetry constrains it 
to keep to the real curve of $\lambda$
shown in figure 1.   The point associated with $z=1$ can be gained from
from $z=-2$ either by going anti-clockwise or clockwise.
Suppose first that $z$ increases monotonically between $-2$ and $1$ as
the Zeeman splitting is varied smoothly from well-split spins, $\Gamma_0(2)$,
to degenerate spins, $\Gamma^0(2)$. 
Figure 2 shows that, as $z$ is increased from $-2$ to $-1$, the
fixed point at $\sigma_*=\frac {1+i} {2}$ of the $\Gamma_0(2)$ flow (figure 3)
follows the semi-circular arc of radius $\frac 1 2$, moving down to the left
until it hits the origin in the $\sigma$-plane, $\sigma_*=0$ when $z=-1$.
It then continues up the imaginary axis, through $\sigma_*=i$
for $z=-\frac 1 2$,  to $\sigma_*=i\infty$ for $z=0$ (near $z=0$ equation (\ref{general-flow}) can be replaced with
\beq
\Sigma_s(\sigma)=\frac{1}{i\pi}
\frac {\tilde z} {(\tilde z \vartheta_2^4 + \vartheta_3^4)},\eeq
where $\tilde z = 1/z$, multiplying $\Sigma_s$ by a constant does not change the
topology of the flow).
The fixed point subsequently
moves down from $\sigma_*=1+i\infty$ to $\sigma_*=1+i$ (a $\Gamma^0(2)$ fixed point,
figure 11) as $z$ increases from
 $0$ to $1$.  While experimental data to date do show a fixed point
on the semi-circle spanning $1$ to $0$ that is to the left of $\frac{1+i}{2}$, \cite{Taiwan2,Taiwan3,Hilkeetal}, it does not seem likely that this sequence of flows can be the correct
one.  A fixed point at $\sigma_*=i$ when $z=-\frac 1 2 $ has never been seen in any Hall sample.
Indeed such fixed points were  identified in
\cite{Gammatheta} as being associated with bosonic pseudo-particles excitations 
(as in the superconductor-insulator transition of \cite{MFisher} for example),
rather the fermionic pseudo-particle excitations of the quantum Hall effect.  
Since no quantum Hall
sample to date has ever exhibited a critical point at $\sigma=i$
we exclude this possibility.

An alternative possibility is that the flow morphs from figure 3 to figure 11
by decreasing $z$ from $-2 $ going through $-\infty$ to 
$+\infty$ to continue down to $+1$.  
The nine plots in figures 3 to 11 show the series of flows for 
\[z=-2,-10,-100,\pm\infty,100,20,10,2 \ \hbox{and}\ 1.\]
The fixed point at $\frac{3+i}{2}$ for samples with well-split
spins moves left and down as the Zeeman splitting is decreased, along 
the semi-circle of radius $\frac 1 2$ centred 
$\sigma=\frac 3 2$, 
until it hits the real axis at $\sigma=1$ (for $|z|=\infty$),
where it merges with the incoming fixed point coming from  
$\frac{1+i}{2}$ on the left.
It then moves vertically upwards to the point $1+i$ when $z=1$,
which is the fixed point for degenerate samples with symmetry $\Gamma^0(2)$.
Every flow in the sequence has $\Gamma(2)$ symmetry, which is enhanced
to $\Gamma_0(2)$ or $\Gamma^0(2)$ at the extreme values $z=-2$ and
$z=1$ respectively.  

In real samples particle-hole symmetry is hardly likely to be an exact symmetry
of the system, there will be deviations from this picture. 
But any deviations will
be small if particle-hole interchange is a reasonably good symmetry:
for example if $\frac{m_p - m_h}{m_p+m_h}$ is small, where $m_p$ is the
particle mass and $m_h$ the hole mass.
In particular the collision of the critical points at $\sigma=1$ ($|z|=\infty$)
seems likely to be an artifact of the mathematical idealisation of
exact particle-hole symmetry, as there is no obvious physical mechanism
governing the merging of two critical points as the Zeeman energy is reduced.
The most plausible scenario here is that the merge is postponed in a real
sample until the Zeeman splitting is reduced to very small values and
the proposed trajectory of a real sample, in which particle-hole interchange
is a good but not exact symmetry, is shown in figure 12.
This is
a small perturbation of the mathematically idealised flows shown in figures 3
to 11 which is still compatible with the proposed symmetries.

\section{Conclusions}

The topology of the temperature flow of conductivities 
in quantum Hall samples is tightly constrained by the law
of corresponding states, expressed in terms of modular transformations
on the complex conductivity (\ref{Complexsigma}).
For the extreme cases of well-split spins and degenerate spins
the critical points in the complex plane are determined by the symmetry.
For intermediate values of the Zeeman splitting modular symmetry does
not determine the position of the critical points but one
can assume that they move around the complex plane in a continuous
manner as the Zeeman splitting is varied.

In the case of samples exhibiting particle-hole symmetry
modular symmetry is particularly powerful, leading to the
statement that the curves in figure 1, and their images under $\Gamma(2)$
modular transformations, will be trajectories of the conductivity
flow as the temperature is varied keeping the magnetic field fixed.
This statement should be true 
for  any Zeeman splitting.  
Zeeman energies which are large enough to give well-split spins
result in critical points at $\sigma_{xy}+i\sigma_{xx}=n+\frac{1 +i}{2}$ 
between integer Hall plateaux 
$\sigma_{xy}=n$ and $n+1$ (figure 3).
Zeeman energies which are so small that the spins are degenerate
give critical points at $\sigma_{xy}+i\sigma_{xx}=2n+1+i$ 
between integer Hall plateaux $\sigma_{xy}=2n$ and $2n+2$ (figure 11).

The form of the flow as the 
Zeeman splitting
is varied from the spin-split to the spin degenerate case is shown
in figures 3 to 11. 
Figure 12 shows
proposed trajectories of two critical points as the Zeeman
energy is reduced in a real sample exhibiting symmetry under 
particle-interchange which is good but not exact.

The analysis here has assumed that the parameter $z$ varies monotonically
as the Zeeman energy is varied, implying that the critical point at $\sigma=\frac{1+i}{2}$
in the transition between $\sigma=1$ and $\sigma=0$
in spin-split samples moves to the right, down towards to $\sigma=1$, as the Zeeman
splitting is reduced.  There is as yet no experimental evidence for such 
behaviour:
indeed in \cite{Taiwan2,Taiwan3,Hilkeetal} a critical point is found on the semi-circle spanning
0 to 1 in the complex conductivity plane which is to the {\it left} of $\sigma=\frac{1+i}{2}$. 
This could be a consequence of a constant re-scaling of $\sigma_{xx}$, \cite{BDspin},
or, perhaps more likely, it may indicate that $z$ is not monotonic.   
It could be that the critical point at $\sigma=\frac{1+i}{2}$ ($z=-2$ in
equation (\ref{general-flow}))
in spin-split samples first starts to move down and to the left ($z>-2$)
and then reverses to retrace its steps back to $z=-2$ before starting to
travel down to the right towards $\sigma=1$ as $z$ decreases below $-2$.
A possible trajectory is shown in figure 13.
While this would be compatible with the experimental data to date
any physical explanation of such a trajectory, which
is certainly allowed by the law of corresponding states combined with
particle-hole symmetry, would go beyond the general predictions
following from these assumptions and would probably require a more specific
microscopic model.
More experimental data
would be welcome in order to determine the true behaviour of the critical 
points.

B.P.D. acknowledges the warm hospitality of Pei-Ming Ho and Chi-Te
Liang during his visit funded by the National Science Council, Taiwan,
R.O.C. (NSC 098-2912-I-002-097). It is a pleasure to thank Pei-Ming Ho
and Chi-Te Liang for discussions.

\vfill\eject

\pagebreak

\centerline{ }
\vtop{
\includegraphics{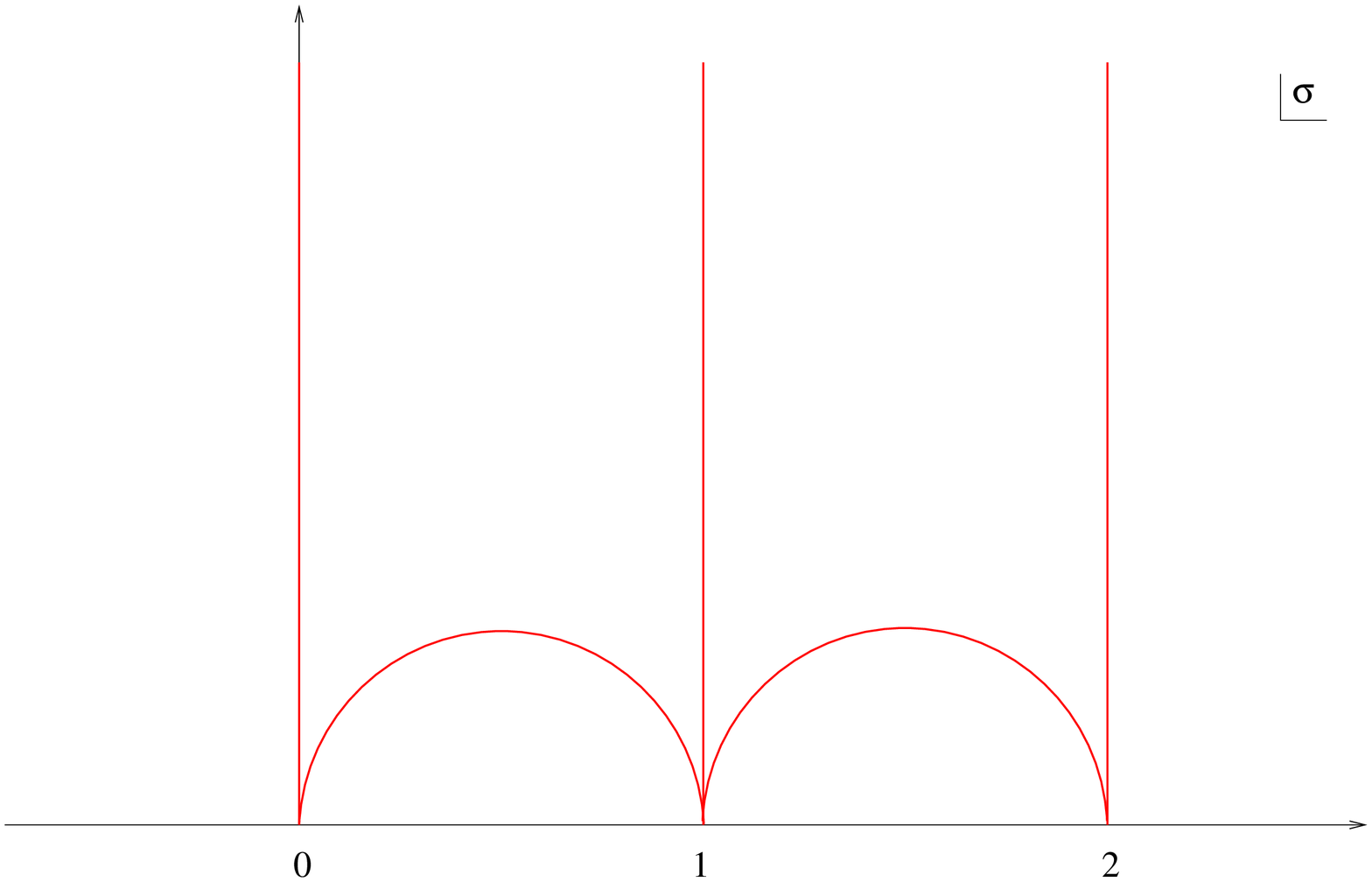}}
\vskip 12cm
\centerline{{\bf Figure 1:} Lines on which the invariant function of  $\Gamma(2)$, $\lambda(\sigma)$ in the text, is real.}

\pagebreak

\centerline{ }
\vtop{
\includegraphics{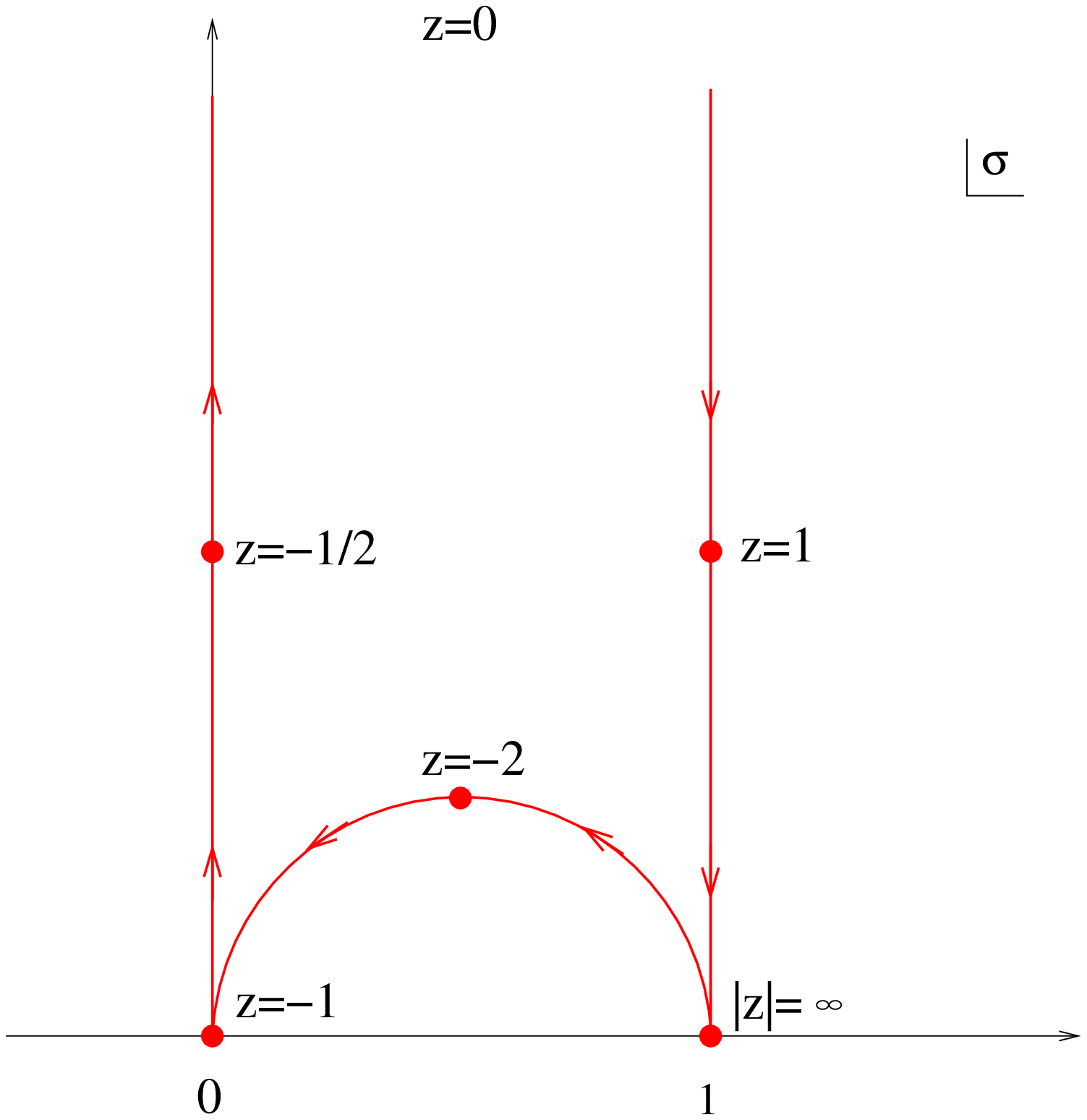}}
\vskip 12cm
\centerline{{\bf Figure 2:} The movement of the $\Gamma_0(2)$ fixed 
$\sigma_*=\frac{1+i}{2}$ as the parameter $z$}
\centerline{is varied away from $-2$. The arrows show
the direction of increasing $z$.} 

\pagebreak

\centerline{ }
\vtop{
\includegraphics{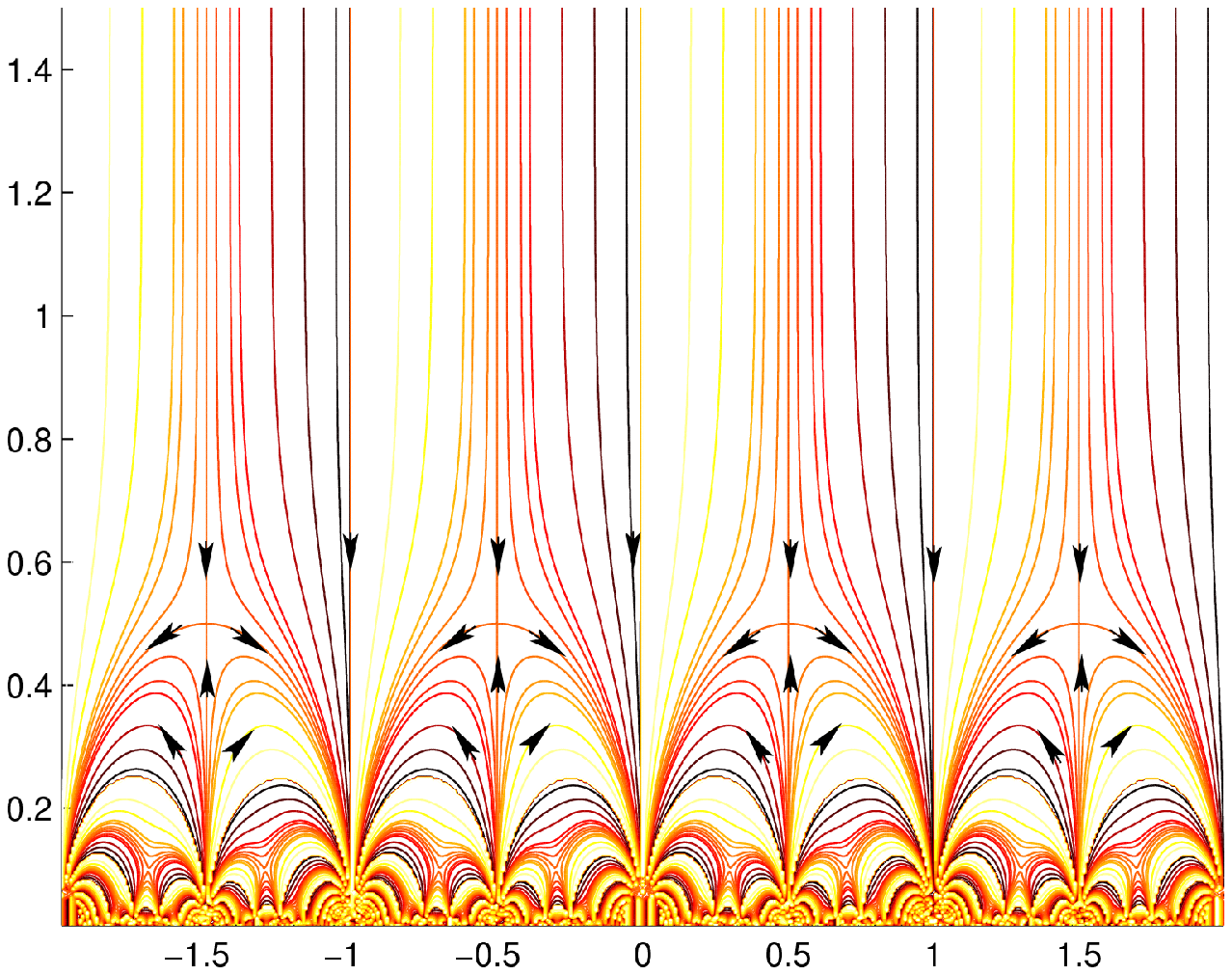}}
\vskip 10cm
\centerline{{\bf Figure 3:} Spins well split, $\Gamma_0(2)$ symmetry.}

\pagebreak

\centerline{ }
\vtop{
\includegraphics{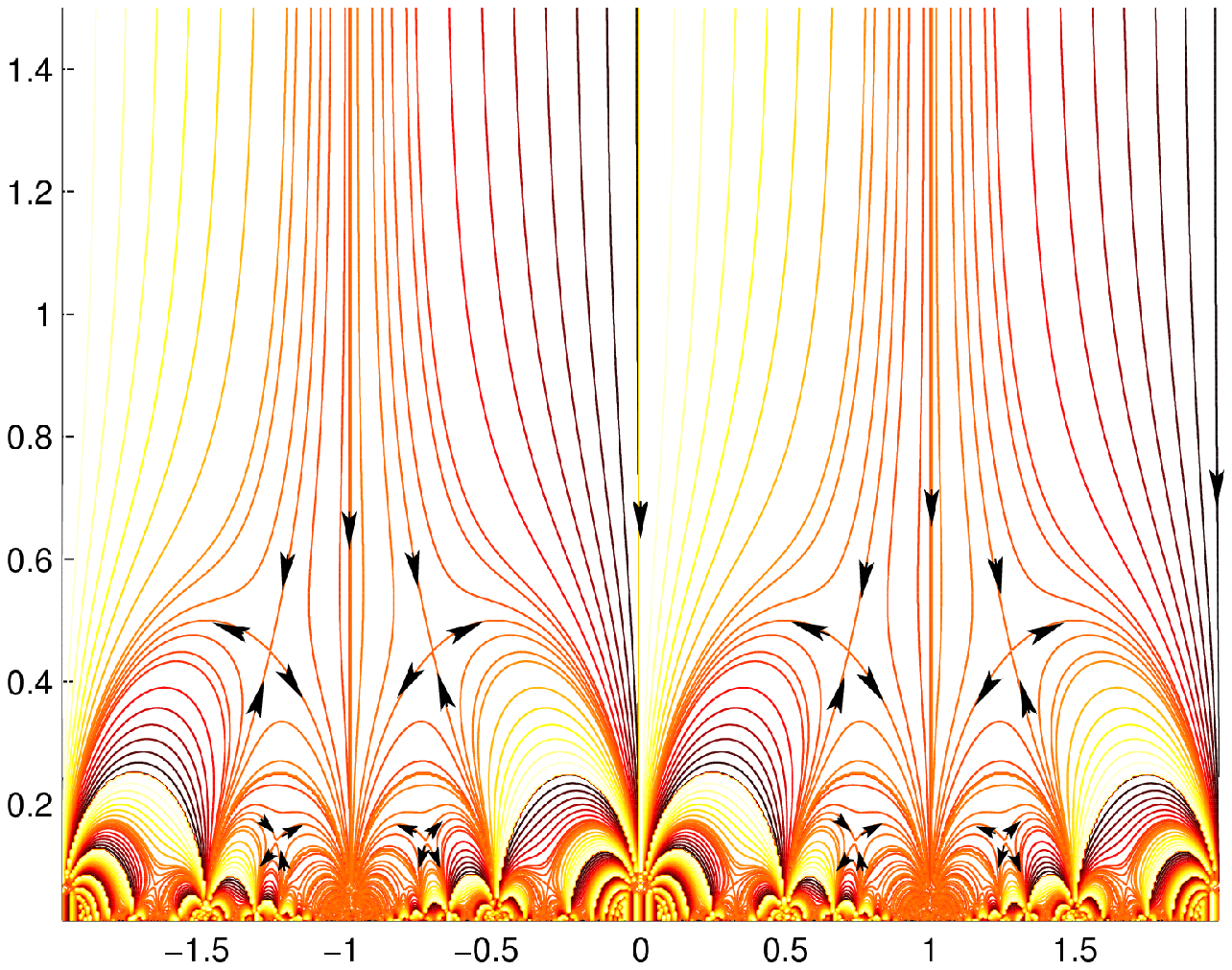}}
\vskip 10cm
\centerline{{\bf Figure 4.}}

\pagebreak
\centerline{ }
\vtop{
\includegraphics{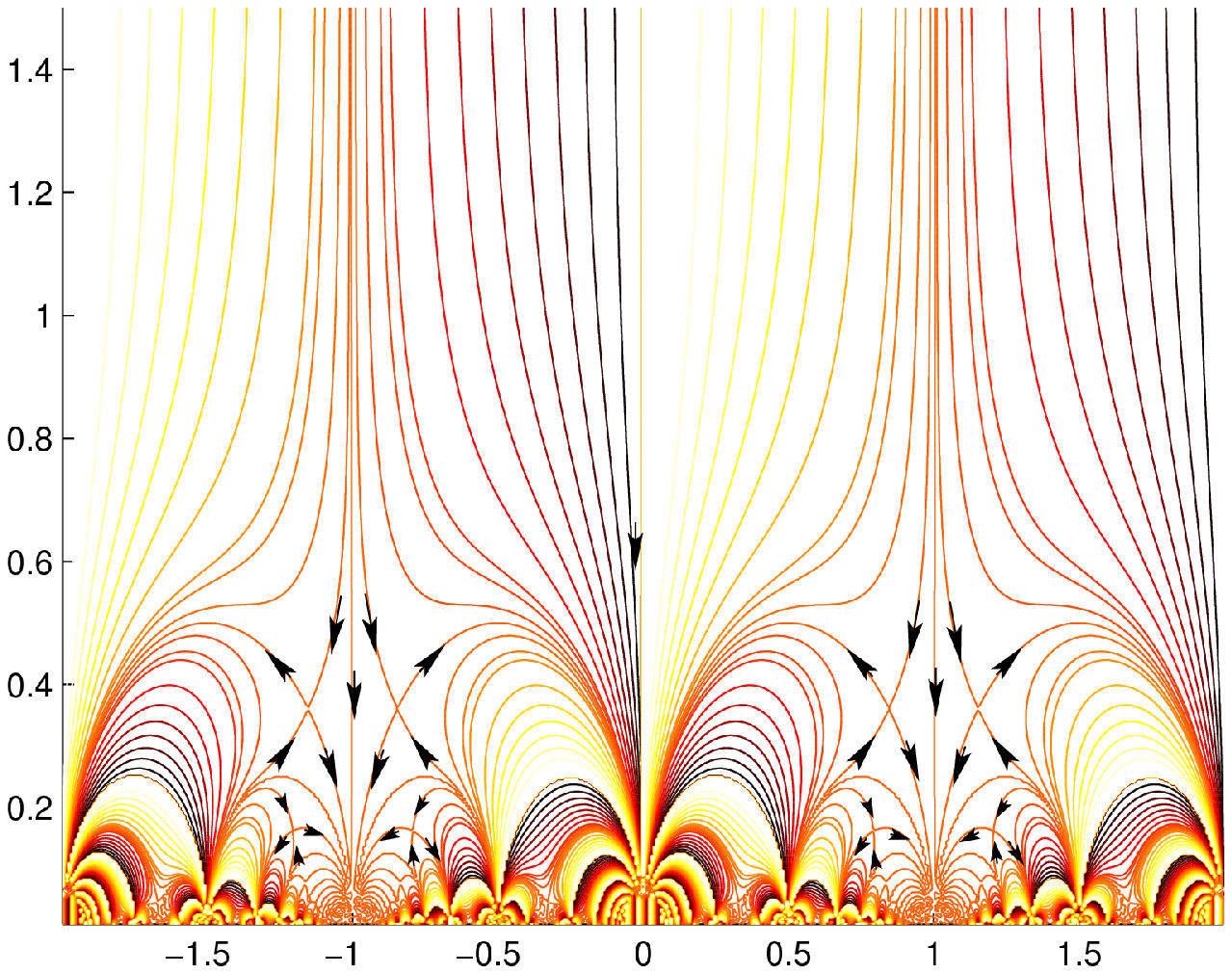}}
\vskip 10cm
\centerline{{\bf Figure 5.} }

\pagebreak

\centerline{ }
\vtop{
\includegraphics{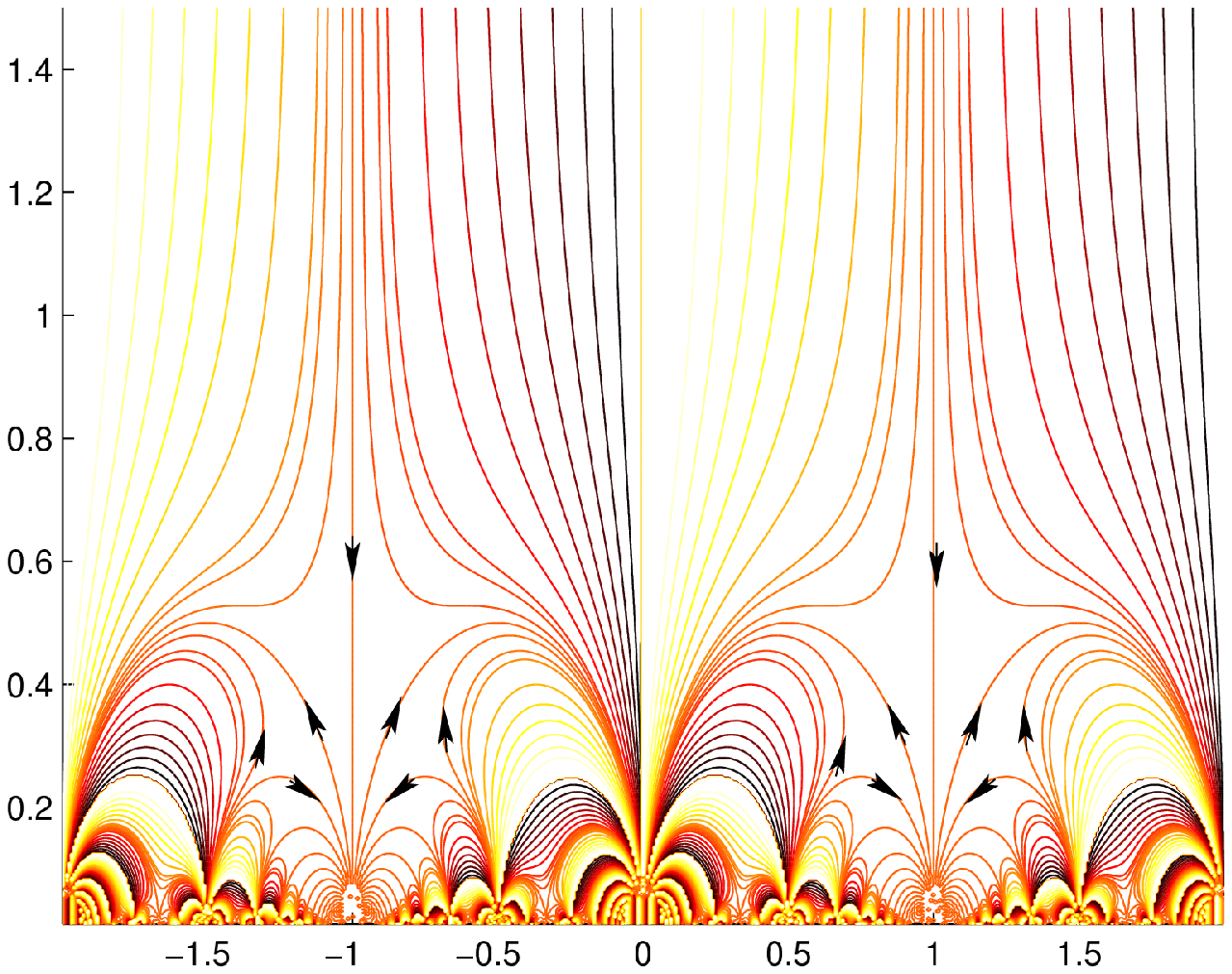}}
\vskip 10cm
\centerline{{\bf Figure 6.} }

\pagebreak

\centerline{ }
\vskip -10cm
\hskip -2.5cm
\vtop{
\includegraphics{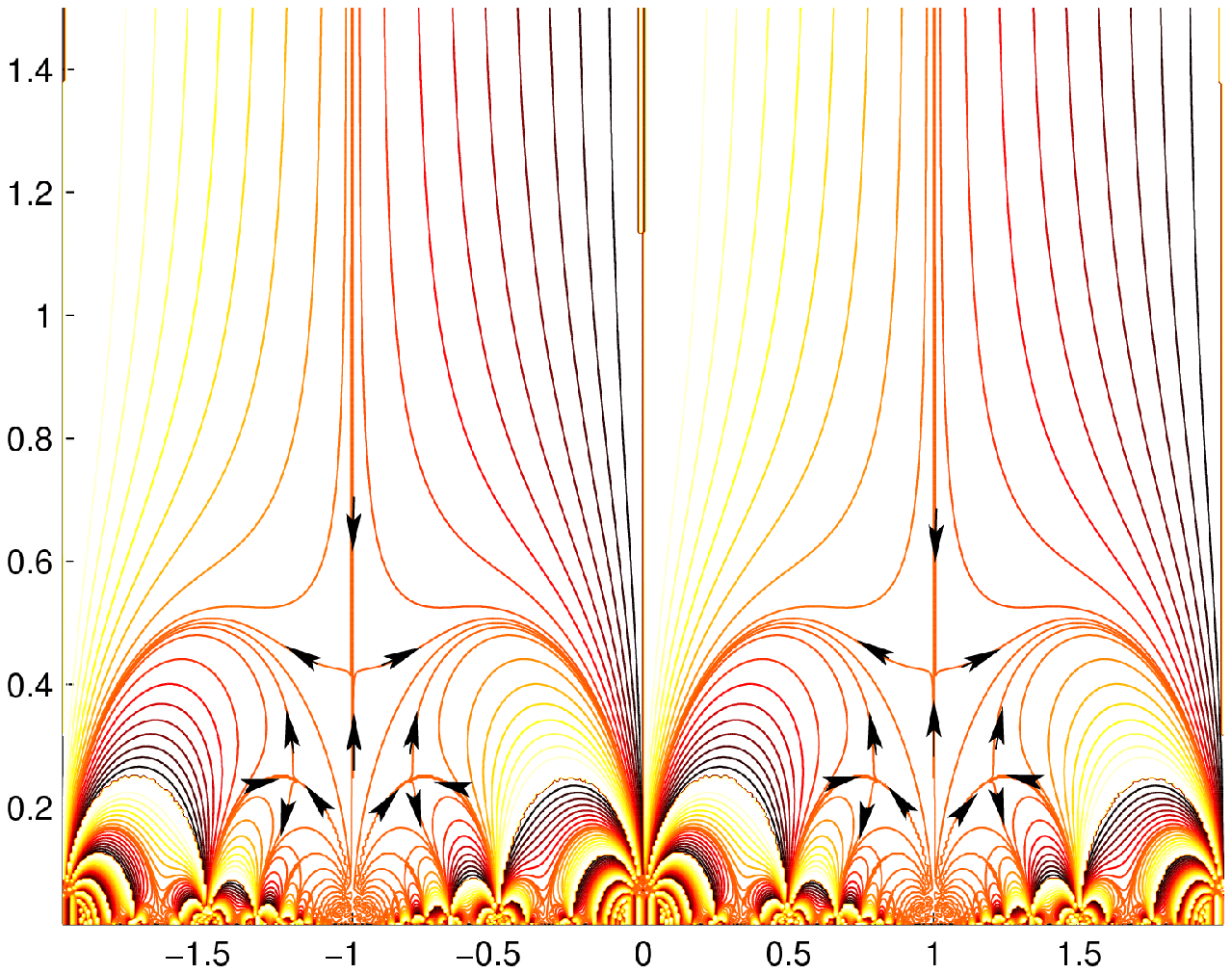}}
\vskip 20cm
\centerline{{\bf Figure 7.} }

\pagebreak

\centerline{ }
\vtop{
\includegraphics{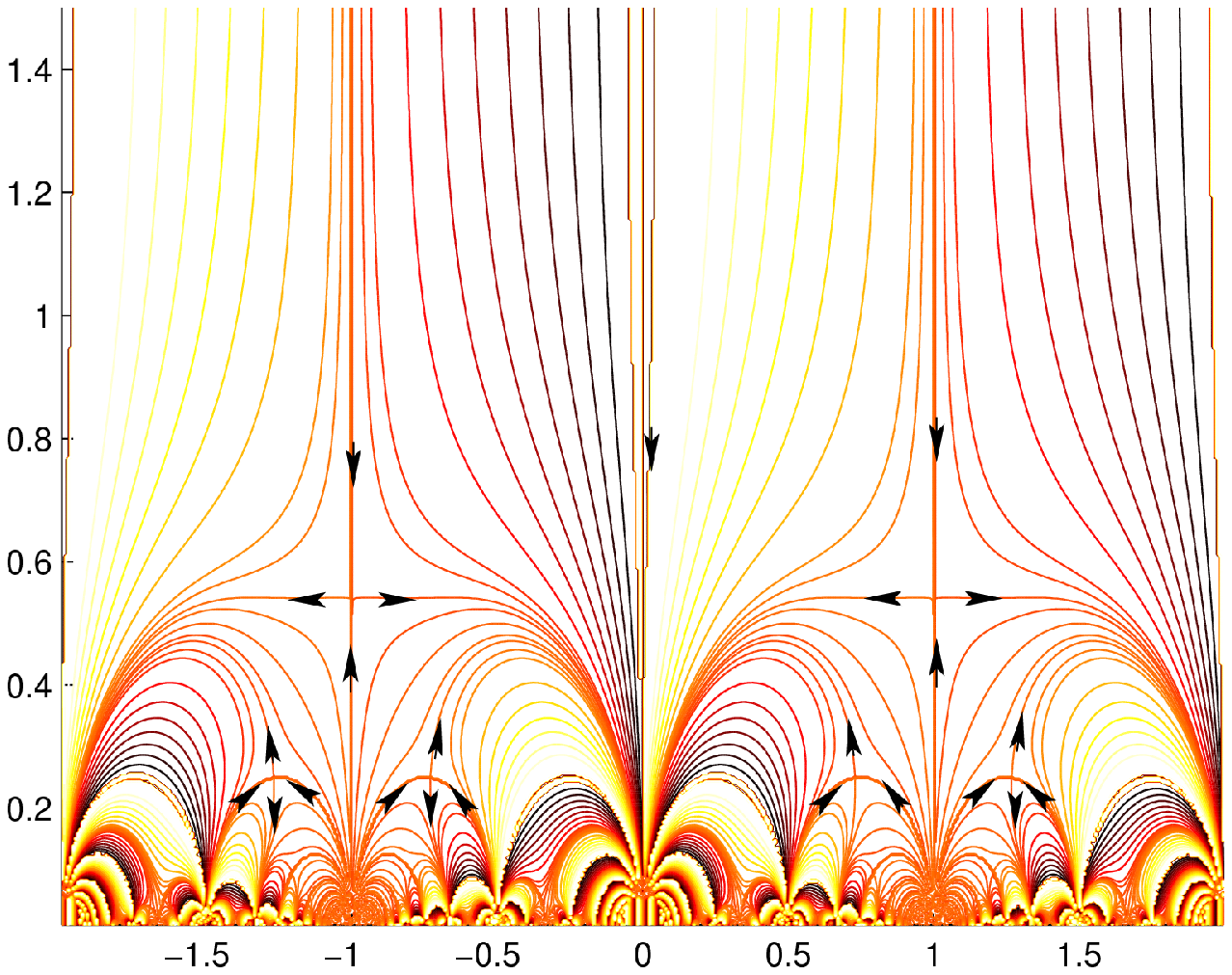}}
\vskip 10cm
\centerline{{\bf Figure 8.} }

\pagebreak

\centerline{ }
\vtop{
\includegraphics{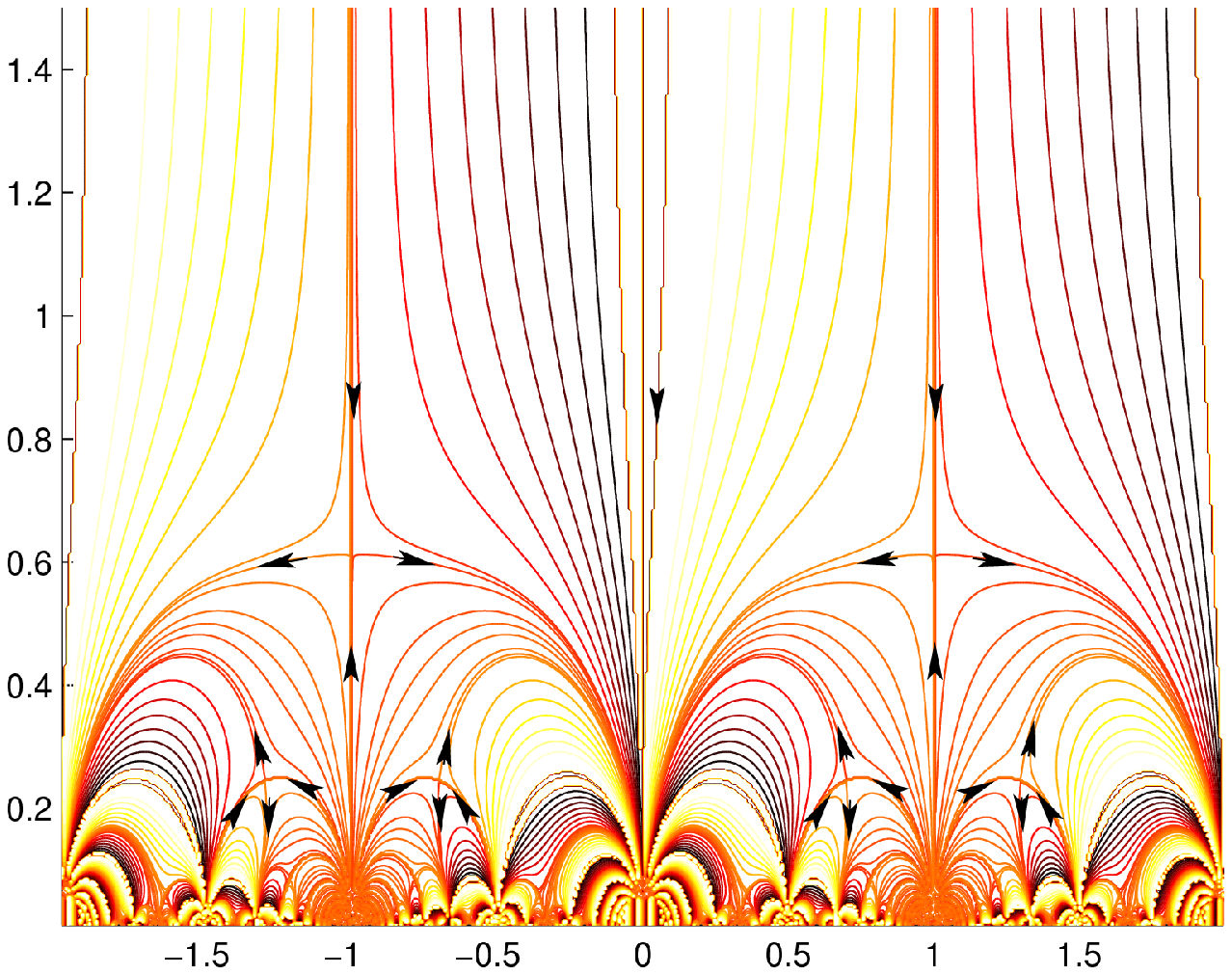}}
\vskip 10cm
\centerline{{\bf Figure 9.}}

\pagebreak

\centerline{ }
\vtop{
\includegraphics{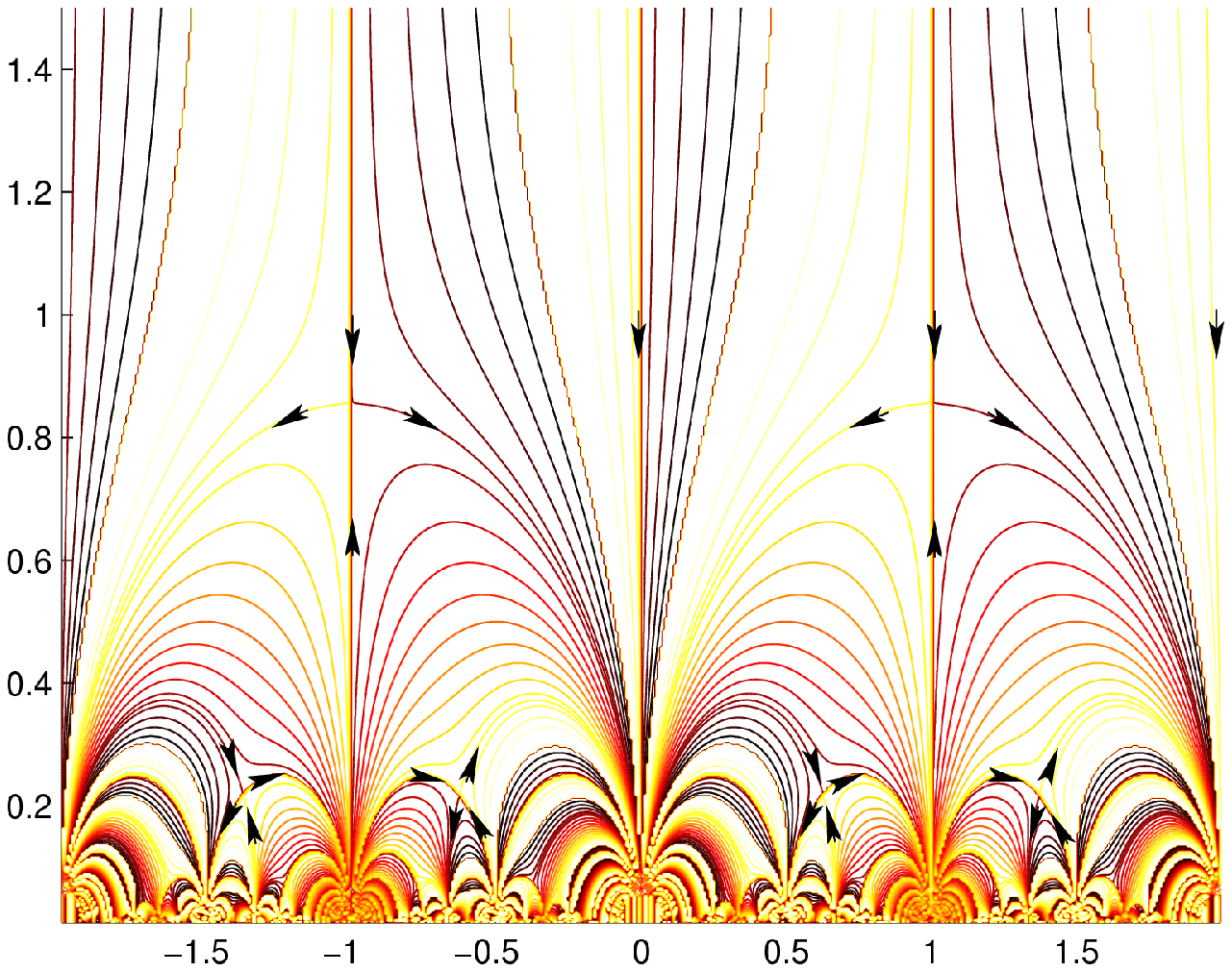}}
\vskip 10cm
\centerline{{\bf Figure 10:} Splitting reduced, $\Gamma(2)$ symmetry.}

\pagebreak

\centerline{ }
\vtop{
\includegraphics{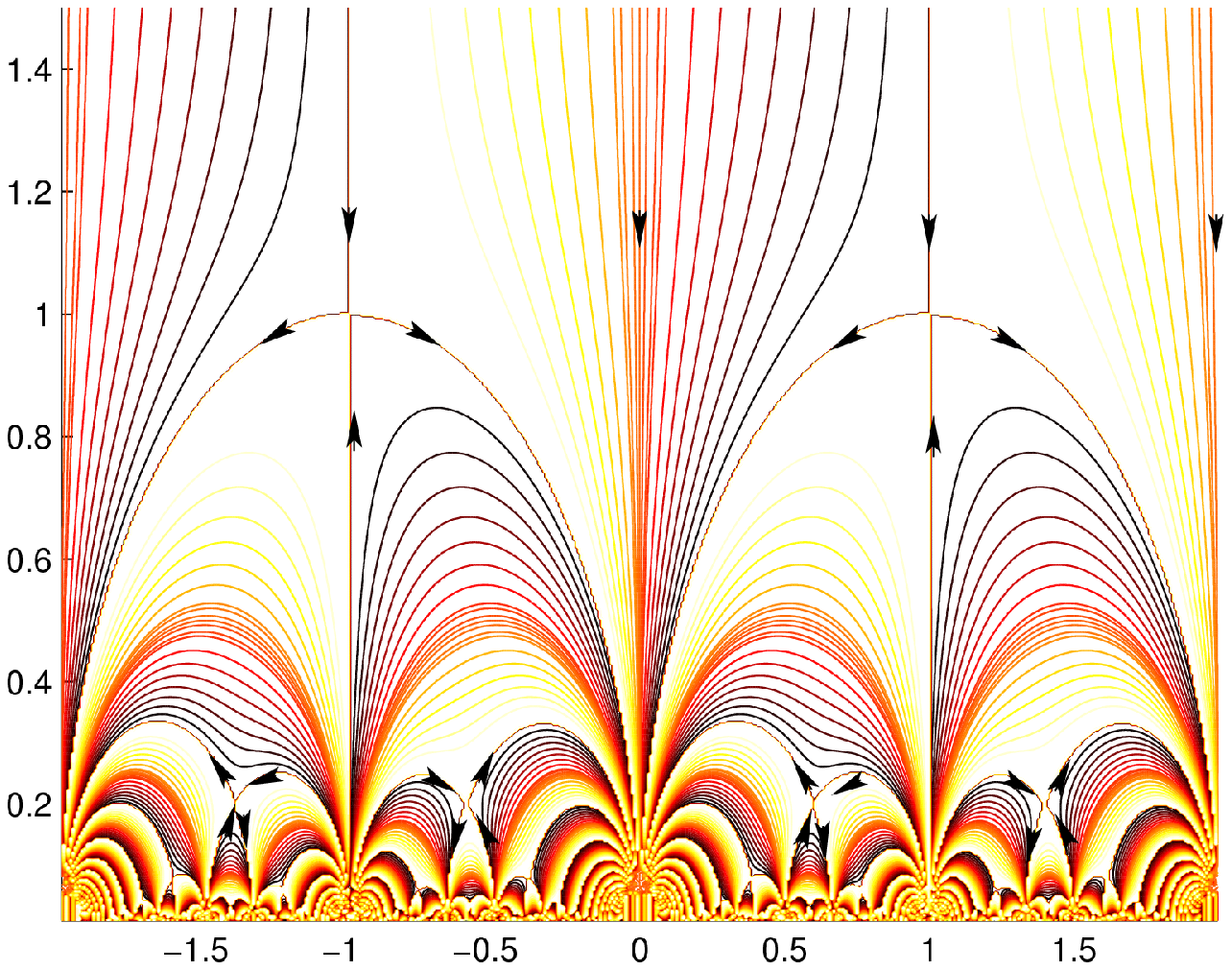}}
\vskip 10cm
\centerline{{\bf Figure 11:} Spins degenerate, $\Gamma^0(2)$ symmetry.}

\pagebreak

\centerline{ }
\vtop{
\includegraphics{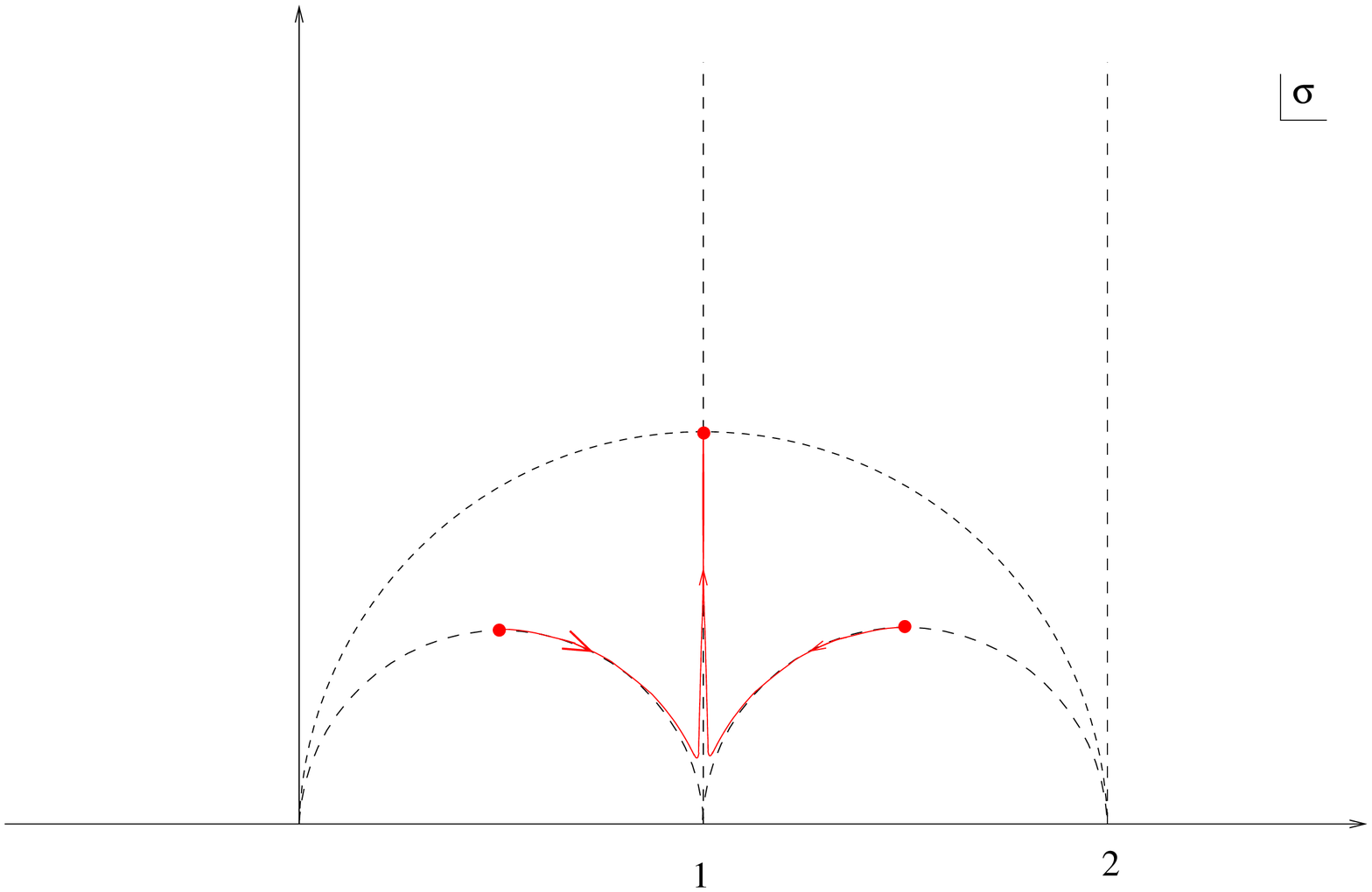}}
\vskip 10cm
\centerline{{\bf Figure 12:} Proposed movement of the fixed point 
$\sigma=\frac {(1+i)} 2$ for non-degenerate spins, $\Gamma_0(2)$,}
\centerline{\hskip 65pt to $\sigma = 1+i$ for spin degenerate samples, $\Gamma^0(2)$, as the
Zeeman energy is reduced.}

\pagebreak

\centerline{ }
\vtop{
\includegraphics{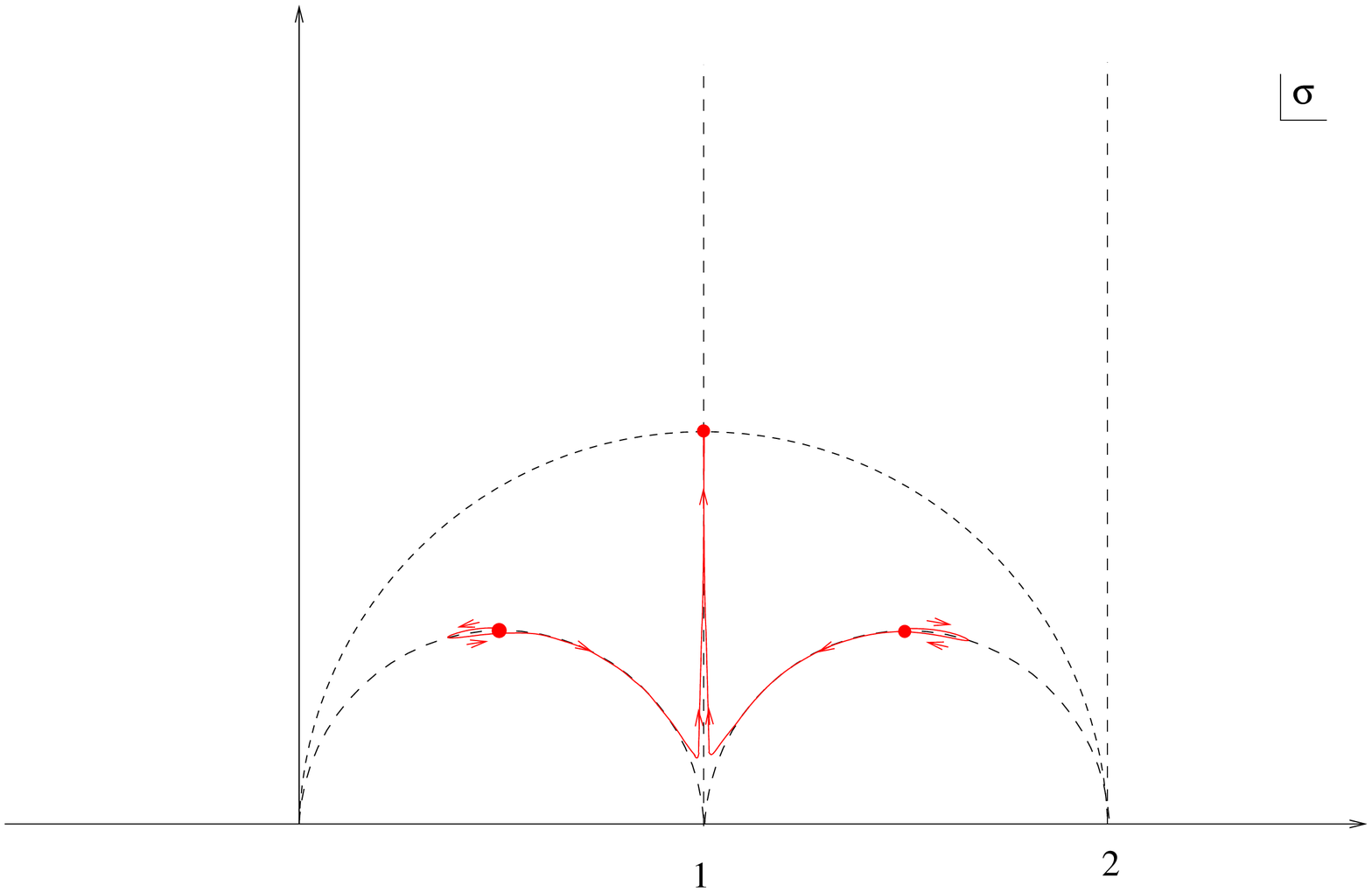}}
\vskip 10cm
\centerline{{\bf Figure 13:} Possible movement of the fixed points
if the mathematical parameter} 
\centerline{$z$ does not vary monotonically with the Zeeman splitting.}

\end{document}